\documentclass[reprint,amsmath,amssymb,aps]{revtex4-1}

\usepackage{graphicx}
\usepackage{epstopdf}

\usepackage[utf8x]{inputenc}
\usepackage{hyperref}
\usepackage{color}
\usepackage[dvipsnames]{xcolor}

\usepackage{footnote}

\begin{document}
 \title{Finite-time performance of a single-ion quantum Otto engine }

\author{Suman Chand}
\altaffiliation[Present  address: ]{Department of Physics, Ko\c{c} University, 34450 Sarıyer, Istanbul, Turkey}
 \email{suman.chand@iitrpr.ac.in}
\author{Shubhrangshu Dasgupta}%
\author{Asoka Biswas}%
\affiliation{%
Department of Physics, Indian Institute of Technology Ropar, Rupnagar, Punjab 140001, India
}%


\begin{abstract}
We study how a quantum heat engine based on a single trapped ion performs in finite time. The always-on thermal environment acts like the hot bath, while the motional degree of freedom of the ion plays the role of the effective cold bath. The hot isochoric stroke is implemented via the interaction of the ion with its hot environment, while a projective measurement of the internal state of the ion is performed as an equivalent to the cold isochoric stroke. The expansion and compression strokes are implemented via suitable change in applied magnetic field. We study in detail how the finite duration of each stroke affects the engine performance.  We show that partial thermalization can in fact enhance the efficiency of the engine, due to the residual coherence, whereas faster expansion and compression strokes increase the inner friction and therefore reduce the efficiency.

\end{abstract}

\maketitle

\section{\label{sec:i}Introduction}

A standard heat engine operates between two heat reservoirs, maintained in thermal equilibrium at different temperatures. The engine operation involves heat absorption from the hotter reservoir and rejection of heat into the colder reservoir, after performing some work, all in a cyclic fashion. For a given set of bath temperatures, the Carnot engine sets an upper bound of the work efficiency that is maximally achievable by any other classical heat engines \cite{Kerson Huang book on Statistical Mechanics}. However, such efficiency is attained only in reversible cycles that ideally run for infinitely long times (quasistatically) and therefore is associated with zero power at output. Running a heat engine for long cycle times is rather impractical. It is important to have more power (work output per cycle) of a realistic engine, even at the cost of lower efficiency. Curzon and Ahlborn indeed obtained an optimized achievable efficiency at maximum finite power \cite{CA efficiency}.

To obtain  finite power, the engines must operate for finite duration per cycle. The study of finite-time performance for heat engines has drawn considerable interest and seen a great deal of progress in recent times \cite{kosloff1984quantum, gevaquantum, rezek2006irreversible, esposito2010finite,  abe2011maximum, wang2012efficiency, wu2006generalized, wang2007performance, abah2012single, wang2012quantum, wang2013efficiency, allahverdyan2013carnot, uzdin2015equivalence, wang2015efficiency, wu2014efficiency, alecce2015quantum, del2014more, beau2016scaling, deng2018superadiabatic, leggio2016otto, zheng2016occurrence, yin2017optimal, shiraishi2017efficiency, jaramillo2016quantum, pezzutto2018out, pietzonka2018universal, scopa2018lindblad, Efficiency of Harmonic Quantum Otto Engines at Maximal Power, peterson2018experimental, camati2019coherence, ccakmak2019spin, Experimental Demonstration of Quantum Effects in the Operation of Microscopic Heat Engines, Efficiency of a Quantum Otto Heat Engine Operating under a Reservoir at Effective Negative Temperatures, Boosting the performance of quantum Otto heat engines-Achieve higher efficiency at maximum power with finite-time quantum Otto cycle, Finite-power performance of quantum heat engines in linear response, Power and efficiency of a thermal engine with a coherent bath, Collective performance of a finite-time quantum Otto cycle, Finite-time performance of a quantum heat engine with a squeezed thermal bath, Non-Markov enhancement of maximum power for quantum thermal machines, Quantum-dot heat engines with irreversible heat transfer, Finite-time quantum Otto engine: Surpassing the quasistatic efficiency due to friction, Quantum Carnot cycle with inner friction, Optimal cycles for low-dissipation heat engines}.  In fact, the finite-time thermodynamics \cite{andresen1984thermodynamics, salamon1980minimum, andresen1996finite, bejan1996entropy} has always been attractive to  researchers. When the cycle time is finite, due to entropy production in the engine cycle, a real heat engine is unable to attain the theoretical limit of the maximum efficiency. This entropy production is the thermodynamic signature of irreversibility \cite{de1984non}. In the classical engine, two types of friction are responsible for irreversibility \cite{camati2019coherence}. One of these is non-conservative frictional force, which is primarily mechanical in nature. On the contrary, the internal friction occurs due to the finite-time operation of the engine.

In the quantum regime, one usually maps the internal friction to the non-commutativity of the driving Hamiltonian at different times
\cite{rezek2010reflections, kosloff2002discrete, feldmann2006quantum, feldmann2012quantumrefrigerator, feldmann2003quantum, rezek2006irreversible, thomas2014friction, plastina2014irreversible}. This involves transition between the instantaneous energy eigenstates. The effect of such friction vanishes when the quantum heat engine operates in the quasistatic limit \cite{rezek2010reflections}. Alternatively, one may also apply the so-called quantum lubrication \cite{feldmann2006quantum} or perform shortcuts to quantum adiabaticity \cite{torrontegui2013shortcuts, del2013shortcuts, del2014more, abah2018performance, ccakmak2019spin}. Several studies attempted to determine how this internal friction affects the performance of a quantum heat engine \cite{rezek2010reflections, kosloff2002discrete, Quantum Carnot cycle with inner friction, feldmann2003quantum, feldmann2006quantum, rezek2006irreversible, feldmann2012quantumrefrigerator, thomas2014friction, plastina2014irreversible, alecce2015quantum, correa2015internal} when it operates in a finite-time cycle. During finite-duration thermodynamic processes, the system never reaches thermal equilibrium with the thermal bath and therefore sustains residual coherence in the energy eigenstate basis \cite{plastina2014irreversible}, which in turn affects the engine performance \cite{camati2019coherence}.

In this paper, we focus on finite-time analysis of a realistic heat engine. The working substance of this engine, a single trapped ion, is treated quantum mechanically. In our earlier studies \cite{Our Paper- Single ion, Our Paper- Two ions, Our Paper- Three ions} we showed how to implement different stages of a quantum Otto cycle using such a system and considered a timescale which is long enough to ensure that one obtains effectively a full thermalization or adiabaticity in the relevant stages. However, to make such a heat engine practically useful (so that one gets finite power, albeit with  reduced efficiency), we must consider its finite-time operation. We look for two different cases: (i) The isochoric heating stage of the cycle is terminated well before the system attains thermal equilibrium with the hot bath. (ii) The adiabatic stages are performed in such a time scale that entropy production occurs. During this stage, the rapid change in energy levels induces non-adiabatic dissipation, i.e., internal friction \cite{rezek2010reflections}. In both cases, the engine performs for a finite time and therefore delivers finite power. We will also investigate how the engine performs for a few cycles.

The main differences between this work and the earlier studies on finite-time heat engines are as follows (i) Here we discuss a realistic quantum system that can actually be used as an engine in a {\it cyclic} manner. (ii) The system-bath interaction is never switched off during the entire cycle, while in the earlier studies, it was assumed that the system interacts with the hot and cold baths, alternatively.  In addition to the reciprocating heat cycles, in which the bath interaction is  alternately switched off and on, there also exist proposals for continuous-cycle heat engines \cite{Kurizki1,Kurizki2}, which is more suitable to realize in the quantum domain \cite{Three-level quantum amplifier as a heat engine- A study in finite-time thermodynamics, The quantum heat engine and heat pump-An irreversible thermodynamic analysis of the three-level amplifier}. Our model of a quantum heat engine genuinely exploits the quantum nature of the system and other quantum operations.

The paper is organized as follows. In Sec. \ref{Sec. II} we describe our single-ion model and discuss how different finite-time strokes of the quantum heat engine can be implemented in such system. In Secs. \ref{Sec. III} and \ref{Discussions}, we explore how the efficiency varies with the time-scale involved. We summarize the paper in Sec. \ref{conclusion}.

\section{Implementation of the heat cycles} \label{Sec. II}
\subsection{Our model}\label{model}
\begin{figure}[h]
\centering
\includegraphics[width=0.40\textwidth]{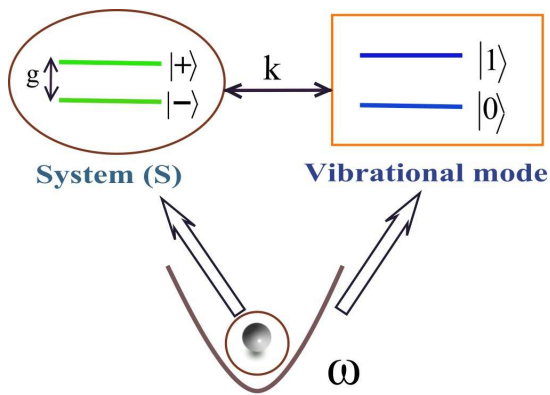}
\caption{Schematic diagram displaying the relevant energy levels of the internal and the vibrational degrees of freedom, relevant to the ionic motion.}
\label{Schematic diagram of the system of single-ion}
\end{figure}
We consider a single trapped ion \cite{Our Paper- Single ion}. In the Lamb-Dicke limit, the ion is confined to its two lowest-lying electronic states $|-\rangle$ and $|+\rangle$ and the lowest-lying vibrational states $|0\rangle$ and $|1\rangle$ (see Fig. \ref{Schematic diagram of the system of single-ion}).  Both the electronic and the vibrational degrees of freedom can therefore be modeled as two-level systems. The following Hamiltonian describes the dynamics of the ion (in units of Planck's constant $\hbar=1$):
\begin{equation} \label{Full Hamiltonain}
H_1(t)=H_S(t)+H_{\rm ph}+H_{\rm int}\;,
\end{equation}
where
\begin{eqnarray} \label{System Hamiltonain}
&&H_S(t)=g\sigma_{x}+B(t)\sigma_{z},\;\;\;H_{\rm ph}=\omega a^{\dagger}a,\\
&&H_{\rm int}=k\left(a^{\dagger}\sigma_{-}+\sigma_{+}a\right).\label{eq:1}
\end{eqnarray}
Here, $H_S$ is the Hamiltonian for the electronic degree of freedom of the ion, $H_{\rm ph}$ represents the energy of the vibrational mode with frequency $\omega$, and  $H_{\rm int}$ defines the interaction between these two modes with the coupling constant $k$. We consider that the electronic states are driven by a local electric field with Rabi frequency $2g$. A time-dependent magnetic field of strength $B(t)$ is applied along the quantization axis. The operators $\sigma_z$ and $\sigma_x$ are usual Pauli spin operators in the basis $(|-\rangle, |+\rangle)$, while the operator $a$ is the annihilation operator operating on the vibrational states $(|0\rangle, |1\rangle)$. We consider the electronic degree of freedom of the ion as the working substance $S$ of the engine, while the vibrational mode plays the role of the effective cold bath. The thermal environment at an ambient temperature $T_H$ is considered as the relevant hot bath.

Traditionally, a bath is a system with many degrees of freedom. Here we consider a finite-dimensional system that interacts with the system (the electronic states) as equivalent to a bath. Any interaction between the two subsystems, in general entangles them and a partial trace of one of these subsystems leads to decoherence in the other. This basic notion of decoherence is used here, where a finite-dimensional system (the two-level vibrational states) is traced partially. That a finite-dimensional system can act as a bath, leading to decoherence in a spin has been shown explicitly in  Ref. \cite{Madam decoherence paper}.   Note that for an average excitation of this mode (vibrational states) $\bar{n}$, one can associate with it an effective temperature using the canonical probability distribution function, as given by $T_L=\hbar\omega/\ln(1/\bar{n}+1)k_B$, where $k_{B}$ is the Boltzmann constant. For $\bar{n}=0.02$, we can have $T_L= 1$ mK (see, for example, \cite{Ion trap paper C. Monroe}), which is a temperature low enough to approximate the vibrational mode as a two-level system, as often routinely done for quantum computing using trapped ions.

\begin{figure}[h]
\centering
\includegraphics[width=0.50\textwidth]{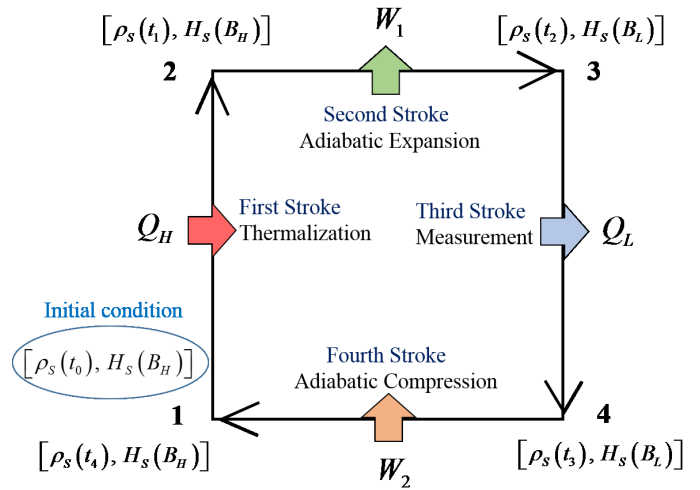}
\caption{Schematic diagram of a quantum Otto cycle.}
\label{Schematic diagram of finite time single-ion QOE}
\end{figure}

We focus on the operation of a quantum Otto cycle in the following. This cycle consists of four stages: two isochoric and two adiabatic stages. Here we show how to use the system $S$ and the two thermal baths as identified above to implement these stages.

\subsection{Stage 1: Isochoric heating}\label{First stroke}  During this stage ($1\rightarrow 2$, Fig. \ref{Schematic diagram of finite time single-ion QOE}), the system $S$ interacts with the hot bath at a temperature $T_{H}$. The magnetic field is kept constant at a value  $B=B_{H}$, which means that the Hamiltonian $H_{S}$ of the system is kept fixed during this stage. If we allow the system-bath interaction for a sufficiently long time, much longer than the thermal relaxation time $t_{\rm{relax}}$ of the system, the system reaches thermal equilibrium with the hot bath.
Due to full thermalization, only the diagonal elements of the system density matrix remain non-zero, when expressed in the eigenstates basis of $H_S$. This means that the system retains no coherence (the off-diagonal elements of the density matrix). On the other hand, the situation is quite different if we consider a partial thermalization, where the time-scale for the system-bath interaction is allowed to be less than or comparable to the thermal relaxation time of the system. In such a case, the coherence in the system does not decay completely. This coherence influences the dynamics of the engine in the next stage of the cycle. We  assume here that the system is initially prepared in its ground state $\left|-\right\rangle $ [the ground state of the system, i.e., the lowest-energy eigenstate of  $H_S(g = 0)=\sigma_{z}$], which means that the system does not exhibit any initial coherence.


The dynamics of the ion under the action of the Hamiltonian $H_{1}(t)$ and a heat reservoir can be described by the Liouville-von Neumann equation  \cite{The Theory of Open Quantum Systems-Book} as 
\begin{eqnarray} \label{Master Equatiom}
\frac{d\rho_{1}}{dt} & = & -\frac{i}{\hbar}\left[H_{1}(t),\rho_{1}(t)\right]\nonumber\\&  &+\left(\bar{n}_{\text{th}}+1\right)\frac{\Gamma}{2}(2\sigma_{-}\rho_{1}\sigma_{+}-\sigma_{+}\sigma_{-}\rho_{1} \nonumber -\rho_{1}\sigma_{+}\sigma_{-})\\&  &+\left(\bar{n}_{\text{th}}\right)\frac{\Gamma}{2}\left(2\sigma_{+} \rho_{1}\sigma_{-}-\sigma_{-}\sigma_{+}\rho_{1}-\rho_{1}\sigma_{-}\sigma_{+}\right),
\end{eqnarray}
where $\Gamma$ is the vacuum decay rate of the electronic states of the ion, $\bar{n}_{\text{th}}$  is average photon numbers of the hot thermal bath at a temperature $T_{H}$, and $\sigma_{-}=|-\rangle\langle +|$ and $\sigma_{+}=|+\rangle\langle -|$ are the usual annihilation and creation operators in the $(|-\rangle,|+\rangle)$ basis,  respectively. In Eq. (\ref{Master Equatiom}) the assumption is the Born-Markov approximation only. The Born approximation is about the weak coupling between the system and reservoir. The Markov approximation is valid when the correlation among the reservoir (hot bath in this case) decreases much faster compared to the relaxation time of the system. Here, of course, we model the reservoir as a collection of infinitely many harmonic oscillators.  Also, Eq. (\ref{Master Equatiom}) does not address the issue of pure dephasing, as we have not considered any information backflow from the hot bath to the system. In this model, the hot bath is coupled to both of the electronic state and vibrational state. However, the proper choice of the thermalization ensures that there is no anomalous heat flow from a hot bath to a cold bath (vibrational state). 
 Now in Eq. (\ref{Master Equatiom}) $\rho_1$ is the joint density matrix of the electronic state and the vibrational mode of the ion and can be written in terms of the joint basis $\left\{ \left|-,0\right\rangle ,\left|-,1\right\rangle \left|+,0\right\rangle \left|+,1\right\rangle \right\} $. We find the evolution of $\rho_1$ and evaluate the density matrix $\rho_S$ of the system $S$ by taking the partial trace over the vibrational state \cite{Our Paper- Single ion}.
We are interested in the heat exchange by the system only. During this stage from $t=t_0$ to $t=t_1$, the heat $Q_{H}$ absorbed by the system can be expressed in terms of the change in average internal energy of the system as
\begin{equation}
 Q_{H}={\rm Tr}\left[\rho_{S}\left(t_{1}\right)H_{S}\left(B_{H}\right)\right]-{\rm Tr}\left[\rho_{S}\left(t_{0}\right)H_{S}\left(B_{H}\right)\right],
\end{equation}
where the initial density matrix $\rho_S(t=t_0)=|-\rangle\langle -|$ evolves into $\rho_S(t=t_1)$ during this stage. Note that the system does not do any work, as the magnetic field is maintained at a constant value.

\subsection{Stage 2: Expansion stage} \label{Second stroke}
During this stage ($2\rightarrow 3$, Fig. \ref{Schematic diagram of finite time single-ion QOE}), the magnetic field is changed from $B_H$ to $B_L$ $\left(B_{L}<B_{H}\right)$ through a duration $\tau$. Usually it is assumed that the heat bath is disconnected from the system. This is true if the evolution takes place within the timescale $\tau\lesssim 1/\gamma$ (where $\gamma$ is the decay rate of the system to the heat bath). This ensures that the heat energy exchanged between the system and environment can be neglected and the driven dynamics becomes a unitarity. In addition, to maintain quantum adiabaticity, one needs to vary the magnetic field slow enough such that $\tau\gg \frac{g}{8}\left|\frac{1}{B_L^2}-\frac{1}{B_H^2}\right| $ \cite{messaiah_book,Our Paper- Single ion}. Clearly for a faster variation, one may attain a nonadiabatic evolution. Here we emphasize that, at the timescale mentioned above, we effectively disconnect the thermal environment from the system, ensuring thermal isolation. This guarantees the thermodynamic adiabaticity. However, during the timescale, it may be possible that internal excitation in the instantaneous energy eigenstates occurs, which indicates that the stroke may be nonadiabatic in the quantum mechanical sense \cite{Nonadiabatic single-qubit quantum Otto engine}.

 We solve the full master equation (\ref{Master Equatiom}), where the Hamiltonian $H_{1}$ is given in  Eq. (\ref{Full Hamiltonain}) and the magnetic field varies linearly with time as given by \cite{Our Paper- Single ion,altintas2014quantum}
\begin{equation}\label{linear}
B(t)=B_H+\frac{B_L-B_H}{\tau}t,
\end{equation}
where $\tau$ is the finite timescale of the change of the magnetic field from $B_H$ to $B_L$ or vice versa. The output of stage 1, the joint state $\rho_{1}\left(t_{1}\right)$, is considered as the initial condition for solving the master equation. This state evolves into $\rho_{1}\left(t_{2}\right)$ a time  $\tau$ later (such that $t_2=t_1+\tau$). As described in Sec. \ref{First stroke}, we can obtain the reduced density matrix $\rho_{S}\left(t_{2}\right)$ of the system $S$, by taking a suitable partial trace. The change in the internal energy of the system during this stage can now be obtained as
\begin{equation} \label{W1}
 W_{1}={\rm Tr}\left[\rho_{S}\left(t_{2}\right)H_{S}\left(B_{L}\right)\right]-{\rm Tr}\left[\rho_{S}\left(t_{1}\right)H_{S}\left(B_{H}\right)\right].
\end{equation}

The above could be considered as the work done by the system if there would be no heat exchange with the bath \cite{Achieve higher efficiency at maximum power with finite-time quantum Otto cycle}. This holds true for the case of a perfect reversible adiabatic process. The time taken to complete the process should ideally be infinite $\left(\tau\rightarrow\infty\right)$, ensuring that the system will remain in the instantaneous eigenstate of the system Hamiltonian. The initial thermal state $\rho_{S}^{\text{th}}(t_{1})$ (when $B=B_H$) remains thermal as $\rho_{S}^{\text{th}}(t_{2})$ at the end of the process, when $B=B_L$
\cite{explain}.  However, for the finite-time nature of the evolution and non-commutativity of the internal and the external part of the Hamiltonian (\ref{System Hamiltonain}), the adiabatic theorem is not expected to hold.  As a result, internal friction arises \cite{rezek2010reflections, kosloff2013quantum, feldmann2000performance, Irreversibility in a unitary finite-rate protocol: the concept of internal friction, ccakmak2017irreversible, plastina2014irreversible}, which has a significant impact on the efficiency of the engine. In this case, the final state of the system deviates from its equilibrium thermal state due to coherence generation in the energy eigenbasis. One can define irreversible work as the extra amount of energy that needs to be done on the system by the driving agent against this friction when the process does not remain reversible. Precisely speaking, if the final state after finite time $\tau$ becomes $\rho_{S}(t_{2})$, instead of the thermal equilibrium state $\rho_{S}^{\text{th}}(t_{2})$, the irreversible work \cite{ccakmak2017irreversible} can be written as
\begin{eqnarray} \label{W1_irreversible}
W_{1}^{\rm ir}&=& W_{1}^{\tau}-W_{1}^{\tau\rightarrow\infty}\nonumber\\
&=&{\rm Tr}\left[\rho_{S}\left(t_{2}\right)H_{S}\left(B_{L}\right)\right]- {\rm Tr}\left[\rho_{S}^{{\rm th}}\left(t_{2}\right)H_{S}\left(B_{L}\right)\right].\nonumber\\
\end{eqnarray}
It is possible to connect this quantity $W_{1}^{\text{ir}}$ to the quantum relative entropy \cite{plastina2014irreversible}. The quantum relative entropy of two density operators $\rho$ and $\sigma$ is defined as, $S\left(\rho||\sigma\right)=k_{B}\left[\rm Tr\left(\rho\ln\rho\right)-\rm Tr\left(\rho\ln\sigma\right)\right]$, which measures the distance between two quantum states. Based on Klein's inequality, the non-negativity of relative entropy ensures that $W_{1}^{\text{ir}}$ is always a non-negative quantity and is expressed as (in units of 
$k_{B}=1$)
%
%

\begin{eqnarray} \label{W_ir(1) for first adiabtic process}
{W^{\text{ir}}_{1}} & = & \scalebox{0.95}[1]{$  T_{H}S\left[\rho_{S}\left(t_{2}\right)||\rho_{S}^{{\rm th}}\left(t_{2}\right)\right] $} \nonumber\\
& = & \scalebox{0.90}[1]{$T_{H}\left({\rm Tr}\left\{ \rho_{S}\left(t_{2}\right){\rm ln}\left[\rho_{S}\left(t_{2}\right)\right]\right\} -{\rm Tr}\left\{ \rho_{S}\left(t_{2}\right){\rm ln}\left[\rho_{S}^{{\rm th}}\left(t_{2}\right)\right]\right\} \right).$} \nonumber\\
\end{eqnarray}


Due to the finite-time Hamiltonian driving, the final system state is out of equilibrium due to the inter-level transitions. Here we examine the quantum relative entropy to investigate how close the nonequilibrium system state is to the corresponding equilibrium state. The quantum relative entropy has a significant thermodynamic interpretation \cite{plastina2014irreversible}. It is related to the inner friction arising due to the finite time adiabatic strokes and the heat exchanges during the cycle's relaxation, which is defined as the difference between the work done on an actual finite-time process and the work done for the quasistatic transformation. Due to the non-negativity of $S\left(\rho\left\Vert \sigma\right.\right)$, the heat exchange during the relaxation process is always positive. This can be interpreted as the additional energy that the driving agent should supply to the system to compensate for the same change due to the fast driving.  This means an extra amount of energy $W^{\rm ir}$ will be stored in the system, making the system state deviate from the equilibrium state and be prepared in a nonequilibrium state. This extra amount of stored energy will be released to the heat bath during the cycle's relaxation process at the beginning of the next isochoric stage \cite{Quantum Carnot cycle with inner friction, Generalized Clausius Inequality for Nonequilibrium Quantum Processes}. Therefore, the net positive work output is decreased [see Eq. (\ref{Eta_ir})].


\subsection{Stage 3: Isochoric cooling} \label{Third stroke}
During this stage ($3\rightarrow 4$, Fig. \ref{Schematic diagram of finite time single-ion QOE}), the system releases $Q_L$ heat to the cold bath and the system Hamiltonian remains fixed at $H_S(B_L)$. Here we employ a projective measurement of the electronic state of the system that post-selects the ground state $|-\rangle$. This essentially is equivalent to a probabilistic cooling of the system \cite{Our Paper- Single ion} and to heat release out of the system. This increases the effective temperature of the vibrational mode, which therefore can be considered as a heat sink. Note that as the probability that the ion is in the ground state $|-\rangle$ is much higher than that for the excited state, the probability of cooling of the system is quite high. Though at the onset of this stage, the system is in an entangled state $\rho_1(t_2)$ in the electronic-vibrational joint basis, the measurement process disentangles them.
After the instantaneous projective measurement, the final state of the system will be
\begin{equation} \label{rho_t3}
\rho_{S}\left(t_{3}\right)=|-\rangle\langle -|.
\end{equation}
So, the heat released from the system to the vibrational mode can be calculated as
\begin{equation}\label{Q_L}
Q_L={\rm Tr}\left[\rho_{S}\left(t_{3}\right)H_{S}\left(B_{L}\right)\right]-{\rm Tr}\left[\rho_{S}\left(t_{2}\right)H_{S}\left(B_{L}\right)\right].
\end{equation}

 The cooling process, as described above, is evidently probabilistic and depends upon the measurement outcome, i.e., post-selection. It may be noted that an alternative way of cooling the system was proposed in Ref. \cite{Kurizki  alternative way of cooling}, which is based on a sequence of nonselective quantum nondemolition measurement of the state of the system $S$, 
in a continuous-cycle scheme. This leads to decoupling of $S$ from the phonon bath, i.e., $\rho_{\rm S,ph} \rightarrow \rho_S\otimes \rho_{\rm ph}$, if the measurement outcomes or, alternatively, the states of the measuring device are not read or averaged out.

\subsection{Stage 4: Compression stage} \label{Fourth stroke}
The magnetic field strength is changed from $B_L$ back to $B_H$, during this stroke ($4\rightarrow 1$, Fig. \ref{Schematic diagram of finite time single-ion QOE}). For infinitely slow variation of the magnetic field, the evolution of the system would be adiabatic. However, as discussed in Sec. \ref{Second stroke}, for finite-time evolution, certain irreversible work $W_{2}^{ir}$ is done, while the system maintains its interaction with both the hot bath and the vibrational mode. To calculate this quantity, we use the initial state of the system plus vibrational mode as the state obtained during stage 3, which can be expressed as $\rho_{1}(t_{3})=\left|-\right\rangle \left\langle -\right|\otimes\rho^{\text{ph}}$, where $\rho^{\text{ph}}$ is the state of the vibrational mode after the projective measurement. The state $\rho_{1}(t_{3})$ evolves into $\rho_{S}(t_{4})$ during this stage, which can be obtained as a solution of Eq. (\ref{Master Equatiom}). The work done on the system during this stage now becomes:
\begin{equation} \label{W2}
W_{2}={\rm Tr}\left[\rho_{S}\left(t_{4}\right)H_{S}\left(B_{H}\right)\right]-{\rm Tr}\left[\rho_{S}\left(t_{3}\right)H_{S}\left(B_{L}\right)\right],
\end{equation}
which includes the contribution from the irreversible work, as given by

\begin{eqnarray} \label{W_ir(2) for first adiabtic process}
{W^{\text{ir}}_{2}} & = & \scalebox{0.95}[1]{$  T_{H}S\left[\rho_{S}\left(t_{4}\right)||\rho_{S}^{{\rm th}}\left(t_{4}\right)\right] $} \nonumber\\
& = & \scalebox{0.90}[1]{$T_{H}\left({\rm Tr}\left\{ \rho_{S}\left(t_{4}\right){\rm ln}\left[\rho_{S}\left(t_{4}\right)\right]\right\} -{\rm Tr}\left\{ \rho_{S}\left(t_{4}\right){\rm ln}\left[\rho_{S}^{{\rm th}}\left(t_{4}\right)\right]\right\} \right).$} \nonumber\\
\end{eqnarray}


 Note that the above protocol of the heat engine is valid for a single run of the engine cycle. However, as in the case of a non-selective
measurement \cite{Talkner non-selective measurement heating}, this would lead to heating of the system when averaged over many cycles. To circumvent this issue, a feedback control \cite{Sagawa feedback control}, namely a local unitary transformation in the system based on the measurement outcome, may be used. Precisely speaking, a $\pi$ pulse (as commonly used in a trapped-ion system \cite{Quantum dynamics of single trapped ions}) needs to be applied to the system, only if it is measured in the state $\left|+\right\rangle $ to return it  to the state $\left|-\right\rangle $ \cite{Elouard QHE Maxweels Demon}. Otherwise, when averaged over many cycles, this would affect the amount of cooling, and thereby reduce the efficiency of the engine, considering the probabilistic nature of projective measurement. However, we have found that the probability that the system will be projected in the excited state is really much lower compared to that for the ground state. This means that for most of the cycles, the system will get cooled.

\begin{figure}[h]
\centering
\includegraphics[width=0.45\textwidth]{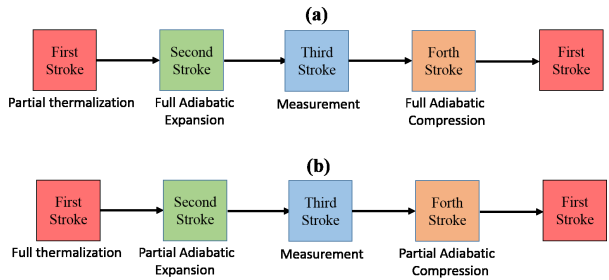}
\caption{Representative diagram of an engine cycle. (a) partial thermalization and fully adiabatic stroke, and (b)  full thermalization and partially adiabatic stroke.}
\label{Schematic diagram of finite time single-ion QOE two cases}
\end{figure}

\section{Efficiency of the heat engine} \label{Sec. III}
In the following, we consider the effect of finite-time driving of the thermalization and adiabatic evolution.

\subsection{Case I: Partial thermalization and full adiabatic stage} \label{Case I: Partial thermalization and full adiabatic stage}

\begin{figure}[h]
\centering
\includegraphics[width=0.40\textwidth]{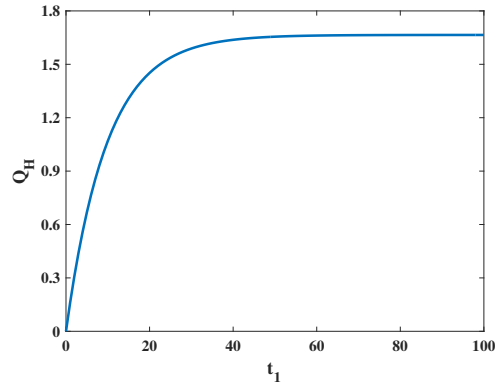}
\caption{Variation of heat exchanged $Q_H$  with the hot bath as a function of the thermalization time $t_{1}$ during first stage. The other parameters are $B_H=10$, $g=0.2$, $k=0.1$, $k_BT_H=10$, $\omega=1$, $\Gamma=0.085$, and  $\bar{n}_{\text{th}}=0.1$. Here we set all parameters to be dimensionless with respect to $\omega$., while $\hbar =1$.}
\label{QH vs t1}
\end{figure}

Here we specifically focus on the effect of partial thermalization \cite{camati2019coherence} on the engine efficiency [see Fig. \ref{Schematic diagram of finite time single-ion QOE two cases}(a) for a detailed protocol]. As discussed in Sec. \ref{First stroke}, the interaction time between the system and hot bath is sufficiently short such that the thermalization is incomplete. However, the adiabatic stages (stage 2 and 4) are considered to run for a suitably long time scale, such that the quantum adiabatic condition is satisf--ied. This  means that there is no irreversible work that otherwise would get wasted from the total work output.
From Fig. \ref{QH vs t1}, it is clear that if the thermalization time $t_{1}$  increases during the first stage, the heat exchange $Q_{H}$ with the hot bath increases as well, and  $Q_{H}$  saturates asymptotically for a long thermalization time. When $Q_{H}$  saturates, it signifies full or complete thermalization, i.e., the system has reached thermal equilibrium with the thermal environment. There is no coherence in the energy eigenbasis in this thermal equilibrium state. However, if we terminate the stroke before saturation, certain coherence is sustained in the system. This coherence can enhance the performance of the engine.

\begin{figure}[h]
\centering
\includegraphics[width=0.40\textwidth]{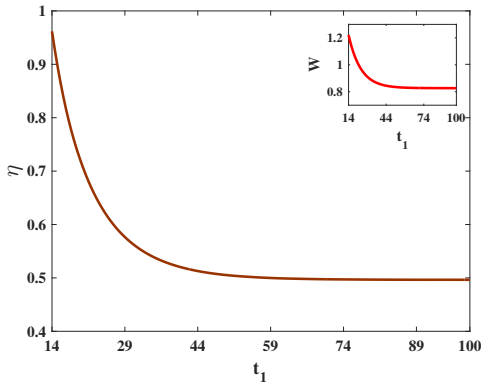}
\caption{Variation of the efficiency $\eta$ as a function of the thermalization time $t_{1}$ during stage 1. Stages 2 and 4 are run in a fully adiabatic fashion. The projective measurement (stage 3) is considered as an instantaneous process and the measurement is performed in the $|-\rangle$ state. During the adiabatic process, the magnetic field is varied from $B_{H}=10$ to  $B_{L}=5$ and vice versa. The other parameters are the same as in Fig. \ref{QH vs t1}. The inset shows variation of work done $W$  by the system as a function of $t_{1}$.}
\label{Partial_thermalisation__Eta vs t1}
\end{figure}

From Fig. \ref{Partial_thermalisation__Eta vs t1} it can observed that for shorter thermalization time $t_{1}$  during  stage 1, the work done by the system $W_1$ increases. This is specifically due to residual coherence in the system during the partial thermalization stage. In such a case, the efficiency of the heat engine, $\eta=\left({\rm work\,output}\right)/\left({\rm heat\,input}\right)=\left(W_{1}+W_{2}\right)/Q_{H}\,$, reaches its maximum value (see Fig. \ref{Partial_thermalisation__Eta vs t1}). Such an enhancement of efficiency can be attributed to the coherence in the system. However, there is a specific value of $t_{1}$ (here $t_{1,{\rm min}}$=14) below which the engine does not operate, as it does not get sufficient heat energy to produce a positive workoutput. When the value of $t_{1}$ is sufficiently large,
the engine efficiency $\eta$ reaches the minimum, which is equal to the maximum efficiency of a perfect single-ion Otto engine \cite{Our Paper- Single ion},
\begin{equation} \label{Single ion Otto engine effciency}
\eta_O=1-\frac{B_{L}}{B_{H}}=0.5 \,.
\end{equation}

\subsection{Case II: Full thermalization and non-adiabatic stage}

Now we investigate the effect of the non-adiabaticity on the thermal efficiency of the engine [see Fig. \ref{Schematic diagram of finite time single-ion QOE two cases}(b) for a detailed protocol]. When the time $\tau$ for the change in the magnetic field is not long, internal excitations between the energy eigenstates of the system can occur, which act as a source of entropy generation in the engine cycle. In this case, we allow the system to thermalize completely with the hot bath during the first stage, while there will be no coherence in the system. However, stages 2 and 4 are run for a duration, when it does not satisfy the quantum adiabatic condition, such that irreversible work is generated.  That is the primary sources of irreversibility in the engine cycle. This is how quantum coherence contributes to irreversibility. Here, we redefine the engine efficiency by considering the effect of irreversible work as

\begin{equation}\label{Eta_ir}
\eta^{\text{ir}}=\frac{(W_{1}+W_{2})-(W_{1}^{\text{ir}}+W_{2}^{\text{ir}})}{Q_{H}} \,.
\end{equation}

\begin{figure}[h]
\centering
\includegraphics[width=0.40\textwidth]{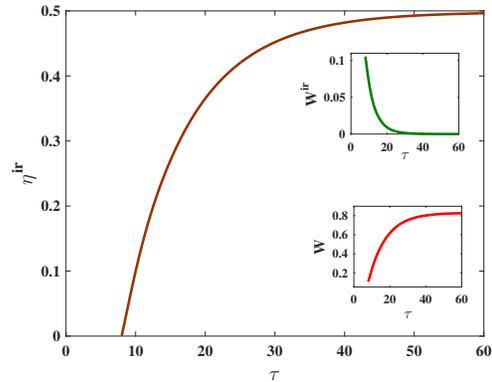}
\caption{Variation of the irreversible efficiency $\eta^{{\rm ir}}$ as a function of the total adiabatic time $\tau$. The thermalization time scale is considered $t_{1}=100$ such that at the end of the first stage, the system attains thermal equilibrium with the hot bath. During stages 2 and 4, the magnetic field is varied from $B_{H}=10$ to  $B_{L}=5$ and vice versa. The others parameters are the same as in Fig. \ref{QH vs t1}. The insets shows the total irreversible work production $W^{{\rm ir}}$ during these two stages and the total work done $W$ by the system during the entire cycle, as functions of $\tau$.}
\label{Partial_Adiabatic__EtaIR vs tau}
\end{figure}

In Figure \ref{Partial_Adiabatic__EtaIR vs tau}, displays the variation of the thermal irreversible efficiency $\eta^{\text{ir}}$  as a function of $\tau$. It is clear that the engine efficiency has strong dependence on $\tau$. For small values of $\tau$, the entropy production in the adiabatic branches greater, which is reflected in the value of total irreversible work $(W^{\text{ir}}=W_{1}^{\text{ir}}+W_{2}^{\text{ir}})$ (see the inset in Fig. \ref{Partial_Adiabatic__EtaIR vs tau}). This shows the presence of internal friction in the system. Also, the total workdone ($W=W_{1}+W_{2}$) by the engine is decreased significantly for the small value of $\tau$. This occurs because, due to the short adiabatic time, the system is unable to follow the evolution of the instantaneous eigenstate of the system Hamiltonian and the occupation probability does not remain the same in this basis.  Further a critical observation in Fig. \ref{Partial_Adiabatic__EtaIR vs tau} reveals that the duration $\tau$ of the expansion and compression stages has a minimum value, (here, $\tau_{{\rm min}}$=8), below which the engine cannot run \cite{Turkpence2019}. This is because, due to smaller $\tau$, the entropy generation is so great that the engine will not be able to produce useful work.  When the adiabatic strokes are performed quasistatically (i.e. $\tau\rightarrow\infty$) such that the adiabaticity is maintained, the engine extracts the maximum possible amount of work. This is because  the quantum adiabatic theorem holds for large $\tau$ and the occupation probability remains the same in the instantaneous eigenstate. As a result, entropy production is minimized, and internal friction is also minimized. The engine efficiency reaches the efficiency of a perfect single-ion Otto engine \cite{Our Paper- Single ion}.

\section{Power of the heat engine} \label{Discussions}
\begin{figure}[h]
\centering
\includegraphics[width=0.40\textwidth]{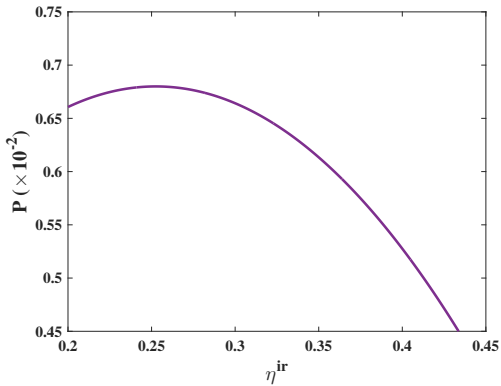}
\caption{Variation of the total power $P$ as a function of the irreversible efficiency $\eta^{{\rm ir}}$ of the engine over the 20 cycles. Here the thermalization time scale is considered as $t_{1}=25$,  such that at the end of the first stage, the system does not reach equilibrium with the hot bath. Also,  stages 2 and 4 are run for a duration $\tau=11$, such that these stages remain non-adiabatic. The other parameters are the same as in Fig.\ref{Partial_Adiabatic__EtaIR vs tau}.}
\label{tbl:many cycles}
\end{figure}

In the preceding section our discussion was limited to a single cycle of the engine. However, from a practical point of view, along with finite time behavior, it is important to study how the engine will behave over many cycles \cite{{watanabe2017quantum}}. In this particular case, the thermalization and adiabatic time are chosen such that thermalization is not complete during stage 1 and stages 2 and 4 do not remain adiabatic.  We start our engine with  the system initially in the ground state $|-\rangle $. So in the next cycle, the initial condition for the stage 1 would change. However, due to the measurement procedure, after just two cycles, all cycles become similar, having the same initial conditions.  A similar observation appears for different initial conditions of the first cycle and different cycle times. We can calculate the average efficiency of the engine  for $n$th cycle as $\eta_{{\rm avg,cy(n)}}^{{\rm ir}}$ equal to the total work for $n$th cycle and the $(n-1)$th cycle divided by the total heat input in the $n$th cycle and the $(n-1)$th cycle, which equal to $ [(W_{{\rm cy(n)}}+W_{{\rm cy(n-1)}})-(W_{{\rm cy(n)}}^{{\rm ir}}+W_{{\rm cy(n-1)}}^{{\rm ir}})]/(Q_{{\rm H,cy(n)}}+Q_{{\rm H,cy(n-1)}})$. Here the power $P$ of the engine in the $n$-th cycle is defined as: $P_{{\rm cy(n)}}=\eta_{{\rm avg,cy(n)}}^{{\rm ir}}/T$, where $T$ is the total time to complete a single engine cycle. From Fig. \ref{tbl:many cycles} we can conclude that the efficiency of the engine at maximum power is 0.25. This indicates that the engine in our model runs with an efficiency, close to the Curzon-Ahlborn efficiency $\eta_{{\rm CA}}=1-\sqrt{B_{L}/B_{H}}=0.292$ \cite{Our Paper- Three ions}, which is attainable at maximum power, by definition.  Note that here we have not considered the  measurement cost \cite{Our Paper- Single ion}. If we include this measurement cost, the actual efficiency $\eta_{{\rm avg}}^{{\rm ir}}$ will decrease below $\eta_{{\rm CA}}$. It may be possible that we can achieve Curzon-Ahlborn efficiency by suitably adjusting thermalization and adiabatic time. Future work may consider the shortcut to adiabaticity to improve the power of the engine  \cite{torrontegui2013shortcuts, del2013shortcuts, del2014more, abah2018performance, ccakmak2019spin}.


 In this model, because a projective measurement has been employed to expedite the process of cooling the system, additionally, a measurement device plays the key role in the process. In that case, this device would indeed be the actual bath, while the vibrational states would just act as an ancillary system. 
 We assumed that this measurement process is instantaneous, and the vibrational mode `acts' as a effective cold bath. 
Note that we have considered the vibrational mode as an effective cold bath, which is confined to its lowest-lying pair of energy levels. So it has a finite heat capacity and its temperature changes during stage $3\rightarrow 4$. However, as the system (i.e., electronic degrees mode) and the effective cold bath (i.e., vibrational mode) thermalize with the hot bath during stage $1\rightarrow 2$ (next cycle), the effective cold bath in our model can be reused for the next cycle of the engine. On the other hand, as we have considered the thermal environment as the hot bath in our model, which has infinite heat capacity, its temperature remains the same during the cycle.

Though we have not presented any interaction model between the system and measuring device, we can choose any arbitrary observable for measurement with the only constraint that the observable must be noncommuting with the system Hamiltonian. It is well known that such an interaction will lead to energy exchange between the system and measuring device. Detection of the state of the measuring device on a suitable basis will be equivalent to the projective measurement of the system's state. Such interaction will also lead to a change in the entropy of both the system and measuring device.  So when the system is projected  from a mixed state to a pure state, the relevant energy cost can be calculated as $M=-k_{B}T_{L}\left(P_{-}\ln P_{-}+P_{+}\ln P_{+}\right)$, where $P_{-}$ ($P_{+}$) are the probabilities of getting the system in the ground state (excited state) after the measurement and $T_{L}$ is the ambient temperature during measurement.  Such an energy cost would refer to erasing (or recording) the measurement outcome, reminiscent of Landauer's erasure principle \cite{Irreversibility and Heat Generation in the Computing Process}. Explicitly,   the projective measurement cost has a maximum value
$M=k_{B}T_{L}\ln2$ \cite{Minimal Energy Cost for Thermodynamic Information Processing_Measurement and Information Erasure}, which corresponds to the erasure of one bit of information (the case of a maximally mixed state). This measurement cost further reduces the effective efficiency of the engine to $\eta_{M}=\left(\rm work\,output\right)/\left({\rm heat\,input}+{\rm measurement\,cost}\right)=\left(W_{1}+W_{2}\right)/\left(Q_{H}+M\right)=\eta/\left(1+\left(M/Q_{H}\right)\right)$. Note that, the denominator $\left(1+M/Q_{H}\right)$ is always positive. For a clear pictorial view about how the efficiency $\eta$ and $\eta_{M}$ (for a maximally mixed state) vary with the work output of the system, we refer to the parametric plot of Fig. 2 in our earlier work \cite{Our Paper- Single ion}. Obviously, the actual power of the engine will also decrease. However, if we reduce our thermalization time below a specific value, the engine does not operate in the physical domain [as discussed before, in Sec. \ref{Case I: Partial thermalization and full adiabatic stage}). This means that $Q_{H}$ always has a minimum positive value and similarly for the case of work output $W$ of the engine (see Fig. \ref{Partial_thermalisation__Eta vs t1}). So with the proper choice of thermalization time, we can operate our engine such that it produces maximal efficiency at a finite power. Note that in recent studies of the thermodynamic aspects of the measurement-based quantum heat engine \cite{Elouard QHE Maxweels Demon, Efficient Quantum Measurement Engines}, the authors also discussed how the measurement cost affects engine efficiency.

 This proposal can be incorporated using the existing trapped-ion technology. As an example, for a single trapped ion \cite{Quantum dynamics of single trapped ions, Measuring the heat exchange of a quantum process}, the trap frequency $\omega$ can be of the order of $2\pi\times10$ MHz, while the time scale of decay of the system, i.e., the electronic excited state can be of the order of $1168$ ms.  Thus the vacuum decay rate $\Gamma$ becomes of the order 0.085$\omega$. In fact, the ion should be cooled in the red sideband limit, where $\Gamma$ (spontaneous decay rate) should be less than $0.1\omega$, which also ensures the Lamb-Dicke regime. Note that the magnitude of $\Gamma$ remains of the same order when the magnetic field changes (if the magnetic field, $B$ is halved, $\Gamma$ becomes $\frac{1}{8}$th). 
 We have checked from the numerical calculation that for moderate changes of $\Gamma$, the efficiency of the engine does not get substantially affected. 
 Moreover,  the average photon numbers of the hot thermal bath $\bar{n}_{{\rm th}}$ should remain much smaller than unity,  which ensures that the average excitation of the vibrational mode is low, such that it can be considered as a two-level system. 
 The temperature $T_{H}$ of the thermal environment can be of the order of a few millidegrees Kelvin \cite{A spin heat engine coupled to a harmonic-oscillator flywheel}.  An external magnetic field gives rise to a Zeeman splitting, typically in the range from $2\pi\times 5$ to $2\pi\times20$ MHz. We can consider a temperature $T_{H}$ such that the average  photon  number  of  the  hot thermal  bath can be $\bar{n}_{\text{th}}=0.1$.  The Lamb-Dicke paramete, proportional to $k$, governs the coupling strength between internal and vibrational states. The typical value of $k$ can be 0.07. The timescale for each engine cycle can be of the  order of a few microseconds. 

\section{CONCLUSION}\label{conclusion}
In summary, we have shown how a single trapped ion can be employed to perform as a finite-time quantum Otto engine. We explored the role of coherence in the engine performance. When the engine operates in full thermalization with partial adiabatic stroke, its performance deteriorates. This is due to the irreversible work (internal friction), which is produced due to fast operation. If we maintain the quantum adiabaticity during engine operation, i.e., if the adiabatic cycle time is long enough, we restore the efficiency of an ideal single-ion Otto engine. On the other hand, when the engine operates in partial thermalization with fully adiabatic strokes, its performance improves. This is due to the residual coherence after the thermalization stroke. Furthermore, when the thermalization time is quite long, i.e., when the system reaches thermal equilibrium with the hot bath at the end of the thermalization stroke, we restore the ideal single-ion Otto engine efficiency. More interestingly, we have shown that if we suitably choose the timescale of thermalization and adiabatic strokes, we can increase the engine efficiency and also produce finite power and   we can attain the efficiency at maximum power, close to the Curzon-Ahlborn bound.


\begin{thebibliography}{50}
\bibitem{explain} The reference to times $t_1$ and $t_2$ here correspond to the values of $B=B_H$ and $B=B_L$, respectively, which are used in the Hamiltonian $H_S$, while calculating $\rho_S^{\rm th}\propto {\rm Tr}_{\rm ph}\exp(-H_1/k_BT_H)$.
\end{thebibliography}
\end{document}